\documentclass[preprint2]{aastex}

\newcommand{\com}[1]{}
\newcommand{\chg}[1]{{#1}}

\slugcomment{To appear in Astrophysical Journal Letters, 564}

\shorttitle{Miyaji et al.}
\shortauthors{Chandra HDF-N Fluctuation Analysis}

\begin{document}


\title{Faint Source Counts from Off-source Fluctuation Analysis 
    on {\it Chandra} Observations of the Hubble Deep Field-North}


\author{Takamitsu Miyaji and Richard E. Griffiths}
\affil{Department of Physics, Carnegie Mellon University,
     5000 Forbes Avenue, Pittsburgh, PA 15213}
\email{miyaji,griffith@astro.phys.cmu.edu}

\begin{abstract}
  We report the results of fluctuation analysis of the off-source 
field from the 1 Million second {\it Chandra} Observation of the 
Hubble Deep Field-North (HDF-N).  The distribution of the counts in 
cells has been compared with the expectations from the 
${\rm Log}\; N - {\rm Log}\; S$ model to constrain the behavior of 
the source number density down to a factor of several lower than the
source-detection limit. Our results show that the number counts 
in the soft band (0.5-2 [keV]) continue to grow down to 
$S_{x}\sim 7\times 10^{-18}$ ${\rm [erg\,s^{-1}\,cm^{-2}]}$, 
possibly suggesting the emergence of a new population and agree
well with a prediction of star forming galaxies by \cite{ptak}.  
For the hard (2-10 keV) band, the fluctuation analysis can loosely 
constrain the source counts fainter than the detection limit and
we found an upper limit of $\la 10000$ $[{\rm srcs\;deg^{-2}}]$ 
at $S_{x}\sim 2\times 10^{-16}$ ${\rm [erg\; s^{-1}\,cm^{-2}]}$. 

\end{abstract}



\section{Introduction}
 
The number count of X-ray sources as a function of flux
(the so called the ${\rm Log}\;N - {\rm Log}\;S$ relation) is 
one of the key constraints for models of the X-ray source 
population. Most of the ``Cosmic X-ray Background'' 
(CXRB) intensity has now been resolved with ``Chandra'' 
\citep{mush,tozzi,br1Ms} to the extent that major uncertainties 
in the fraction of the CXRB which have been resolved into individual 
sources, i.e, how much of it remains to be explained, lie in the 
absolute intensity of the CXRB and field-to-field fluctuations due 
to cosmic variance.  In terms of the origin of the CXRB, the faintest 
sources in the {\it Chandra} Deep Fields are becoming less 
interesting. However, constraints on number counts at the  
faintest possible fluxes provide a new view on the nature 
and evolution  of the X-ray emitting sources.

 Fluctuation analysis is a strong tool for constraining the 
source counts below the source detection limit. 
This technique has been successfully applied to data from previous 
missions \citep{has93,geo93,gendreau,yamashita,perri} and the 
source counts inferred from these analyses have turned out to 
be consistent with those from resolved sources in deeper observations. 
The Chandra observations of the Hubble Deep Field North (HDF-N) 
and Chandra Deep Field-South (CDF-S), with an exposure of about 1 Ms each, 
are the deepest X-ray imaging data obtained so far. Fluctuation 
analyses on these fields are an essential step in pushing the limit 
of the source counts to even fainter fluxes. 

 In this letter, we report our initial results of the fluctuation
analysis of the 1 Ms of the {\it Chandra} observation of HDF-N.
In this first analysis, we have assumed that the sky fluctuation 
solely comes from unresolved point sources (AGNs and galaxies). An 
analysis including the CDF-S, further HDF-N exposures,
and possible effects of extended X-ray emission 
will be reported in a future paper (paper II, in preparation). In this 
letter, we use $H_{0}=65$ ${\rm [km\,s^{-1}\,Mpc^{-1}]}$, 
$\Omega_{\rm m}=0.3$, and $\Omega_{\rm \Lambda}=0.7$ unless 
otherwise noted.
  
\section{Data Preparation}

 The {\it Chandra} X-ray Observatory (CXO) was used to observe
the HDF-N field with its ACIS-I detectors. The total exposure time 
attained as of Spring 2001 has been 975 [ks] over
12 observations and the data, which have been aligned
to an astrometric accuracy of $\sim 0\farcs5$, have been fully  
archived\footnote{http://asc.harvard.edu/udocs/ao2-cdf-download.html}.  
We have made use of the Level 1 merged event list from the archive to 
create images for our analysis. The charge transfer inefficiency (CTI), 
has been corrected according to procedures in \citet{cti}. After this 
correction, we have filtered events to remove hot pixels, columns and 
flaring events flagged in the archived eventlist.

 In order to reduce the particle background as much as possible,
we have used the {\em restricted} grades \citep{br1Ms}, i.e., 
{\em fltgrade}=0,64 for the {\em soft} (0.5-2 keV) band and 
{\em fltgrade}=0,2,8,16,64 for the {\em hard} (2-8 keV) band
respectively. Using these restricted grade sets, we removed 46\% and 
25\% of the off-source events (non X-ray background dominated), while 
removing only 15\% and 13\% of source events for the {\it soft}
and {\it hard} bands respectively compared with the standard 
selection ({\em grade}=0,2,3,4,6). There are two locations where CXO 
pointings are concentrated in the 1 Ms of the HDF-N data, one near  
$(\alpha,\delta)=(189\fdg 2680,62\fdg 2150)$ and the other near 
$(\alpha,\delta)=(189\fdg 1476,62\fdg 2437)$. For the fluctuation 
analysis, we 
have chosen a region which is an {\em intersection} of two 
5$\farcm$2-radius circles centered  at these two locations. We have also 
excluded the locations corresponding to the gaps between CCDs by 
imposing a total exposure value of at least 800 [ks].  The cell size 
used for making histograms of the detected counts was 
$4\arcsec\times4\arcsec$, which was chosen to match the largest 
point-spread function (PSF) in this area. The region contains neither 
of the two extended sources discussed by \citet{br1Ms}. The flux 
(before Galactic absorption)-to-countrate conversion factors have been 
calculated using PIMMS 
\footnote{http://heasarc.gsfc.nasa.gov/Tools/w3pimms.html} 
assuming a $\Gamma=1.4$ power-law with a Galactic absorption of 
$N_{\rm H}=1.6\;10^{20}$ ${\rm [cm^{-2}]}$ corrected for the loss
of events for using the restricted grade set.

\subsection{Point Source Detection and Removal}
\label{sec:srcs}

We have used the {\em wavdetect} utility distributed as a part of the 
Ciao 2.1 package\footnote{http://asc.harvard.edu/ciao/} to find 
sources to be masked out from the image to be analyzed. 
We have used a false detection probability threshold of $1\; 10^{-6}$ 
and wavelet scales of $1,\sqrt{2},2,2\sqrt{2},4,4\sqrt{2},8$ pixels on 
$0\farcs 98$-pixel images.  The detection completeness has been 
extensively investigated using running the same source detection
procedures on simulated images (Sect.\ref{sec:sim}). 
The sources with wavelet-detected counts larger than 12(18) counts,
corresponding to $S_{\rm x16}=0.7$(6.) in 0,5-2 (2-10) [keV]
(here and hereafter, $S_{\rm x16}$ represents an X-ray flux 
measured in  $10^{-16}{\rm [erg\,s^{-1}\,cm^{-2}]}$) have been
masked out from the field for the soft (hard) band. The simulation
shows that less than one source per field fails to be detected above
these thresholds. We take conservatively large radii
for the source exclusion region, ranging from 6$\arcsec$ to 
$\sim 18\arcsec$, depending on the wavelet-detected source
size and counts. After the source removal, the remaining area
for the fluctuation analysis was 7848(8043) pixels or 34.9(35.7) 
arcmin$^2$ for the soft(hard) band. 

 Sources detected in the same procedure over a larger area of the sky 
(within 6$\farcm$3 from both of the pointing centers; 78 [arcmin$^2$]) 
are used to give the ``resolved source''  constraints on the 
fluctuation analysis discussed in Sect. \ref{sec:stat} and to calculate 
the resolved source ${\rm Log}\; N - {\rm Log}\; S$ shown in
Sect. \ref{sec:res}
    
\section{Off-source Fluctuation Analysis}
\subsection{Overall Procedure}
 Using the off-source map generated above, we have searched 
for the behavior of  
${\rm Log}\; N-{\rm Log}\; S$ below the
source detection limit which is consistent with the off-source
fluctuation. For the model, a broken power-law form for the 
cumulative source count has been assumed:
\begin{eqnarray}
     N(>S)= 
     \left\{ \begin{array}{r@{\quad:\quad}l}
       N_0 (S/S_{\rm max})^{-\gamma_1} & 
		(S_{\rm b}\leq S \leq S_{\rm max}) \\
       N(>S_{\rm b}) (S/S_{\rm b})^{-\gamma_2} & 		
		(S_{\rm min}\leq S<S_{\rm b}) \\
		\end{array}
     \right.
\label{eq:bknpow}
\end{eqnarray}
, where $S_{\rm min}$ and $S_{\rm max}$ are the minimum and maximum
fluxes between which we would like to constrain the behavior. 
We set $S_{\rm max}$ as the limiting flux of the complete source
detection and $S_{\rm min}$ as the flux corrsponding to $\sim 1$
count in the image. 
 Confidence level searches in the parameter space have been made 
for two cases, each with two free parameters:(1) Single power-law 
with free parameters ($N_{\rm 0}$,$\gamma_1$), and (2) Broken power-law 
with free parameters ($S_{\rm b}$\,$\gamma_2$), where the 
$N_{\rm 0}$ and $\gamma_1$ values fixed to the best-fit of the
single power-law case.  The second power-law component helps 
determine the practical sensitivity limit of this analysis, 
\chg{following the approach by \citet{has93}}. For each point 
in the parameter space, we have run Monte-Carlo simulations and compared
simulated and observed histograms of the number of pixels
as a function of the number of detected counts (the so-called
$P(D)$ diagram) as detailed in the following
subsections. By running the simulations over the parameter 
space, we have determined the subspace where the model was 
accepted at a 90\% confidence level.  For the accepted subspace,
we have calculated the minimum and maximum $N(>S)$ values at each 
flux between $S_{\rm min}$ and $S_{\rm max}$ to find final constraints 
on the ${\rm Log}\;N - {\rm Log}\;S$ behavior. The resuling constraints
are shown and discussed in Sect. \ref{sec:res} 

\subsection{The Image Simulation}\label{sec:sim}

 The image simulation has been made using simulation software 
developed by one of the authors (TM). It has been designed to 
meet the needs of massive Monte-Carlo simulations of a summed image
from multiple observations with offset pointings. For each 
observation, it uses a separate exposure map and a different pointing 
position, from which the off-axis angle is calculated.

 An image simulation runs as follows. Firstly, random point sources are 
generated based on an input ${\rm Log}\;N - {\rm Log}\;S$ model. 
For a given input source  (with a sky position and a physical flux), 
the simulator generates Poisson-deviated number of events based 
on the {\em maximum} exposure value of {\em each} observation. 
The off-axis angle is calculated from the pointing direction
of the observation and the simulator spatially spreads the 
generated events based on the off-axis angle dependent PSF.  For each 
of the PSF-deviated  event, a random number between 0 and 1 is
generated. The event is rejected if this random number is larger 
than the ratio of the exposure value {\em at the point} and 
the {\em maximum} exposure. This two-step procedure assures the correct 
treatment near the CCD edges and bad columns. The above steps are 
repeated for all the observations and all the generated sources.
Background events are then added to the image in such a way that the 
total count of the sources and the background is the same as  
that of the real data. The background events have been distributed 
assuming that they are dominated by the particle background and 
it is uniform over the active pixels of the CCDs and not affected by 
the vignetting of the CXO telescope. 

\chg{We have forced the total count of the simulated image 
to be equal to the real data, instead of also considering the 
Poisson deviation of the residual background count. This is because the 
underlying background rate is not an interesting parameter here and 
in order to assess  the confidence range of interesting parameters 
(the ${\rm Log}\;N-{\rm Log}\;S$ parameters in this case) for given data, 
uninteresting parameters should be adjusted to give the best fit. In our 
case, forcing the total counts to be equal automatically 
makes this adjustment.}

\subsection{Statistics}\label{sec:stat}
   
 The probability distribution of the $P(D)$ histogram is more 
complicated than the simple Poissonian or Gaussian, 
involving source and photon fluctuations with correlated errors. 
Thus using analytical formulae of the probability distributions
of common statistical measures (e.g. the $\chi^2$ distribution),
gives inaccurate results. Thus we use Monte-Carlo simulations to calculate 
the probability distributions. With Monte-Carlo simulations, the probability 
that a model is correct can be estimated using any reasonable estimator 
which indicates the deviation between the model and the data. 
In this work, we have used a modification of the \citet{cash} estimator 
by Caster, as used in the XSPEC 11.1 \citep{xspec}, the change of which 
usually follows the $\chi^2$ probability distribution:     
%\begin{eqnarray}
%   C = 2 \sum_i [N_{{\rm mdl},i}-N_{{\rm dat},i}+ \nonumber\\
%    N_{{\rm dat}_i} (\ln \,N_{{\rm dat},i}-\ln \,N_{{\rm mdl},i})],
%\label{eq:c}
%\end{eqnarray}
\begin{equation}
   C=2\sum_i [N_{{\rm mdl}_i}-N_{{\rm dat}_i}+ 
    N_{{\rm dat}_i} (\ln \,N_{{\rm dat}_i}-\ln \,N_{{\rm mdl}_i})],
\label{eq:c}
\end{equation}

where $N_{{\rm dat}_i}$ and $N_{{\rm mdl}_i}$ are the observed and 
model-predicted numbers of pixels having $i$ counts ($0\leq i\leq n$)
or the number of resolved sources ($i=n+1$, observed or predicted; 
see below) respectively. The sum is over $2\leq i \leq n+1$.   
We have excluded $i= 0,1$ from the sum because these two bins are 
not independent from the others (the total numbers of pixels and 
events are fixed). As an additional constraint, a term ($i=n+1$) where 
the number of {\em resolved} sources above the defined flux limit 
is $N_{{\rm dat},n+1}$ and its model-predicted value $N_{{\rm mdl},n+1}$ 
has been included in Eq. \ref{eq:c}.
This term constrains the ${\rm Log}\;N - {\rm Log}\;S$ at the brighter 
end of the fluctuation analysis.
 
 For a given model, 1000 simulations have been made and the mean of the 
simulated results has been taken as $N_{{\rm mdl},i}$ \chg{as \citet{has93}
did. However, instead of using the $\chi^2$ distribution, 
we have used the distribution of 1000 $C$ values between this  
$N_{{\rm mdl},i}$ and the 1000 {\em simulated} $N_{{\rm dat},i}$ 
histograms.} If $>10$\% of the simulated $C$ values are greater than 
that between the model and the {\em real data}, the model has been 
accepted at a 90\% confidence level. 

\section{Results and Discussion} \label{sec:res}

\begin{figure*}
%\begin{inlinefigure}
\centerline{\includegraphics[width=0.7\textwidth]{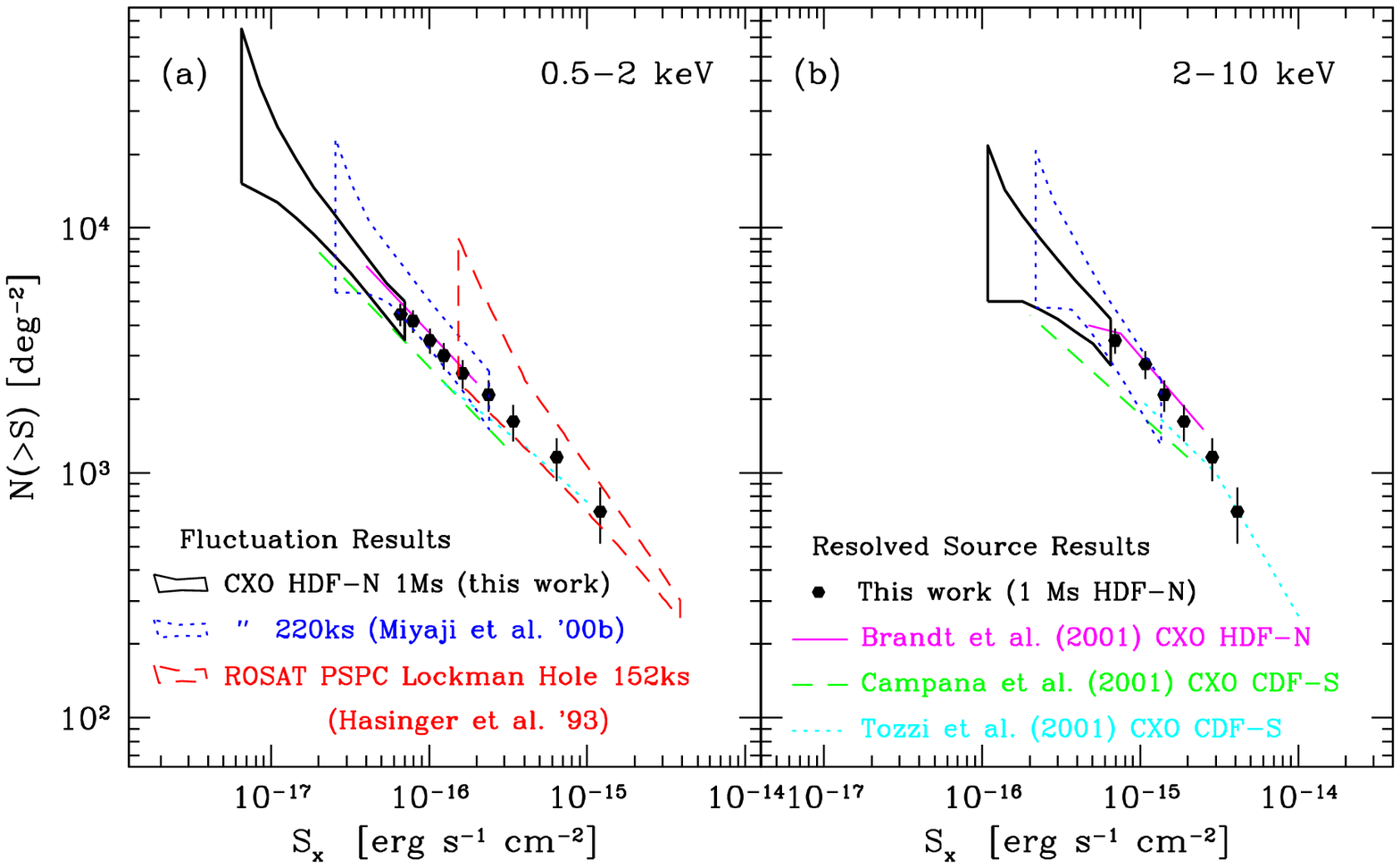}}
%\plotone{lnls_fluct1Ms_col.eps}
\figcaption[f1.eps]{
%\caption{ 
Fluctuation and resolved source  ${\rm Log}\;N-{\rm Log}\;S$  relations 
in the 0.5-2 and 2-10 keV bands 
are shown with those from a number of other works as labeled. 
The solid circles are from the resolved-sources (Sect. \ref{sec:srcs}).   
The solid horns are from the fluctuation analysis.   
Results from a number of recent publications are overplotted as labeled.
For a verification, the results of our previously reported 
analysis for the 220 ks of data \citep{miy_aas} are also
shown. They correctly predicted  the results of the 
{\em resolved} source number counts of the much deeper 1 Ms 
data. \label{fig:lnls_fluct}}
%\end{inlinefigure}
\end{figure*}

 The resulting constraints (90\% confidence range) on the 
${\rm Log}\; N - {\rm Log}\; S$ behavior in the  
0.5-2 [keV] and 2-10 [keV] bands (the latter have been
scaled from 2-8 [keV] assuming a $\Gamma=1.4$ power-law spectrum) 
obtained from our analysis are shown in Fig. \ref{fig:lnls_fluct} 
along with a number of recent results from the literature in the 
flux regime of interest. The resolved source counts (Sect.\ref{sec:srcs}) 
have also been plotted with 1$\sigma$ error bars.
\footnote{ASCII tables of the results can be found in the src
distribution of this preprint and    
at http://astrophysics.phys.cmu.edu/$\sim$deepxray/tables} 
These are consistent with those from \cite{br1Ms}, 
who used sources from a smaller part of the same general area to avoid 
incompleteness. Our resolved source counts are from a larger area and 
are plotted down to fluxes which were not affected by incompleteness 
as determined from the simulations. The fluctuation results have been 
plotted down to a flux where the number counts have been reasonably 
constrained. We have set the faintest sensitivity limit at the flux 
where the difference between the upper and lower bounds is a factor of 
three. We note that we could determine the sensitivity limit 
in this way by exploring the parameter space in a broken power-law form 
(Eq. \ref{eq:bknpow}), and not just the single power-law model. 

 For verification, we have also overplotted the results of our previous 
analysis, which utilized essentially the same procedure, when only 220 [ks] 
of CXO data were available on this field \citep{miy_aas}. 
As 1 Ms of data have been obtained, fainter sources have now been 
resolved.  The resolved source counts in the 1 Ms of observation agree 
well with the 220 ks fluctuation results. 

 Fig. \ref{fig:lnls_fluct} shows that the deep resolved-source counts
in CDF-S by \cite{cam01}, whose faint limit was attained by full 
corrections for incompleteness and the Eddington bias, are somewhat 
below the lower limit of our fluctuation results in both energy bands. 
Comparing a number of results from the literature 
\citep{tozzi,cam01,br1Ms,mush}, the X-ray source counts from CDF-S, 
as well as those of the Lockman Hole observed with XMM-Newton\citep{xmmlh1}, 
are in general lower than those of the HDF-N region by $\la 20-40\%$, 
depending on the work cited and flux range.  This probably reflects 
actual cosmic variance in the sky with some minor contribution from 
different calibrations and source detection techniques.   

\begin{figure}[t]
%\begin{inlinefigure}
\centerline{\includegraphics[width=\hsize]{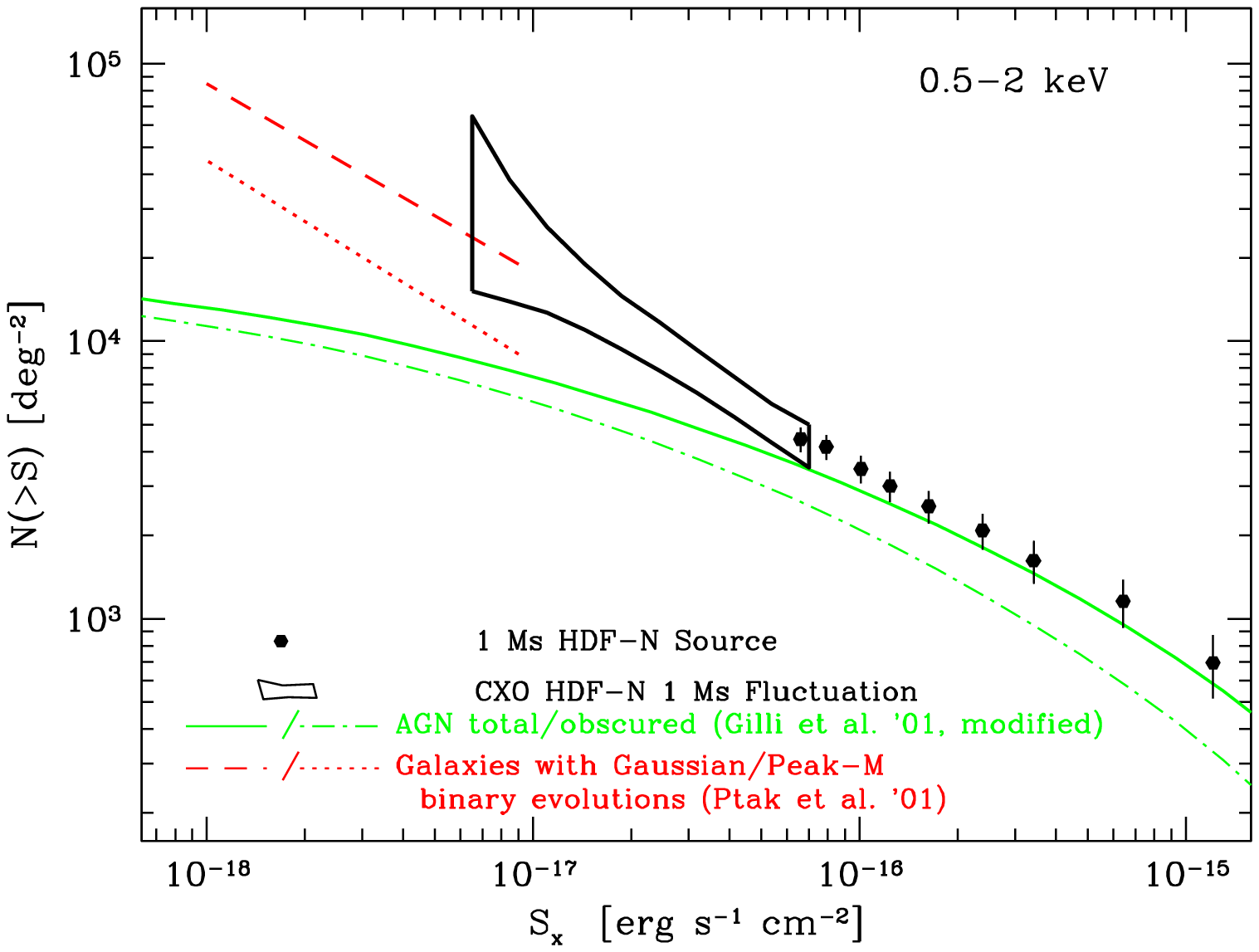}}
%\centerline{\psfig{file=lnlssft_mdl0_col.eps,width=0.9\hsize}}
%\plotone{lnlssft_mdl0_col.eps}
\figcaption[f2.eps]{
%\caption{
Derived ${\rm Log}\;N-{\rm Log}\;S$ relation 
in the soft band is compared with AGN and galaxy number count models.
Predictions from the population systhesis model (B) by
\citet{gilli} for the total AGN population and the obscured 
($N_{\rm H}>10^{22}$ $[cm^{-2}]$). Predicted number counts of galaxies 
using the cosmic star-formation history and two modes (Gaussian and Peak-M) 
of evolution of X-ray binaries \citep{ptak} are also plotted. If 
mode B of \citet{gilli} represents the correct behavior of absorbed
and unabsorbed AGNs, the fluctuation constraints suggest the 
emergence of an additional population, probably from galaxies. 
\label{fig:lnls_mdl}}
%\end{inlinefigure}
\end{figure}

 The most significant result from our analysis is that the 
0.5-2 [keV] ${\rm Log}\;N - {\rm Log}\;S$ continues to grow down to  
$S_{\rm x16}\sim 0.07$. The 2-10 [keV]  ${\rm Log}\;N - {\rm Log}\;S$ 
fluctuation results are consistent with both starting to saturate and 
continuing to grow below $S_{\rm x16}\la 7$, while some loose constraints 
can be made (e.g. $N(>S)< 10000$ at $S_{\rm x16}= 2$).   
Here we focus our discussion on the implication of our 
results in the soft band. From the deepest {\it ROSAT} 
surveys \citep{rds6}, the X-ray sources at $S_{\rm x16}\ga 20.$ 
turns out to be dominated by AGNs, while at fainter fluxes, 
difficulties in complete optical identifications of the faint sources 
detected in {\it CXO/XMM-Newton} ($S_{\rm x16}\ga 1$) leave us with 
tantalizing situations in the nature and redshifts of these sources. There 
still exist large uncertainties in the behavior of the soft X-ray luminosity 
function (SXLF) of AGNs at moderately low luminosities 
($L_{\rm x}\sim 10^{42}-10^{43}$ ${\rm [erg\,s^{-1}]}$) at $z\sim 0.5-2$. 
\citet{miy_xlf1,miy_xlf2} constructed an SXLF of AGNs 
($L_{\rm x}\ga 10^{41.5} {\rm [erg\,s^{-1}]}$) using {\it ROSAT} surveys 
with a very high degree of completeness in spectroscopic identifications 
down to $S_{\rm x}\sim 20$. They discussed two extremes of possible 
extrapolations of the SXLF behavior in the framework of the 
luminosity-dependent density evolution (LDDE) picture. One of them 
has rapidly dropping density-evolution rates as luminosities become 
lower ($L_{\rm x}\la 10^{44}$ ${\rm [erg\,s^{-1}]}$) (LDDE1) and the 
other has only a moderate drop in evolution rate in such a way that 
AGNs alone can make $\sim 90\%$ of the soft X-ray background (LDDE2). 
The LDDE2 model overpredicts the number count by a factor of two at  
$S_{\rm x16}= 1$ and thus can be considered to be rejected by the new data. 
We have overplotted the 0.5-2 keV ${\rm Log}\;N - {\rm Log}\;S$ prediction 
from the AGN population synthesis model composed of unabsorbed and absorbed 
AGNs based on the LDDE1 model by \citet{gilli} (model B). The plotted 
model have been slightly modified from the original as described in 
\citet{rosati}. The curves for the total and absorbed ($N_{\rm H}>22$  
${\rm [cm^{-2}]}$) AGN populations are plotted.  Fig.\ref{fig:lnls_mdl} 
shows that the AGN counts drop below $S_{\rm x16}= 1$ even if the emergence 
of the obscured AGN population is taken into account,  while the source counts 
from the fluctuation analysis continue to grow.  If their model B closely 
represents the true behavior of the SXLF of the AGN population, we are 
probably seeing the emergence of a new population of faint X-ray sources.  
This excess may be contributed by AGNs with low intrinsic luminosities 
(at $L_{\rm x}\la 10^{41.5}$ ${\rm [erg\,s^{-1}]}$) and/or X-rays from 
star-formation activities through supernova remnants and low-/high- mass 
X-ray binaries. In view of this, we have also overplotted 
a model prediction of X-ray number counts based on the cosmic star-formation 
rate \citep{ptak} for the two models of binary evolution 
(Gaussian and Peak-M) from \citet{ghosh}.  Within the uncertainties in 
the AGN source counts and model construction, these predictions gave 
approximately the number counts from our fluctuation analysis. This 
is also consistent with the results of stacking analysis of bright 
galaxies by \citet{br500ks}.

 Finally we describe a number of caveats to be considered in
interpreting our analysis. The possisble inhomogeneities in the CCD 
quantum efficiencies and particle backgrounds, and the existence of 
low-surface brightness diffuse sources (unresolved galaxy groups 
and possible intergalactic medium) would all work in the directions of 
increasing the apparent number counts inferred from the fluctuation analysis. 
The excellent agreement between the 220 ks fluctuation results and 1 Ms 
source counts shows that at least the first effect is negligible. 
Furthermore, summing data from 12 observations have further smoothed 
out the first two effects. However, actual diffuse structure in the sky 
may come into effect at this very faint level. This aspect will be discussed 
in paper II.       

\acknowledgments
This work has been made using data from the {\it Chandra} Data
Archive. The authors acknowledge the support from the 
NASA LTSA Grant NAG-10875 (TM) and the CMU subcontract 
of NASA grant NAS8-38252 to Penn State U. (REG). The authors 
thank the members of the ACIS team for their help with the analysis 
and in particular, L. Townsley  for her effort on creating and 
maintaining the CTI corrector, Bob Warwick, Xavier Barcons and 
the referee, G\"unther Hasinger for useful discussions and comments,
and R. Gilli for sending the machine-readable tables of the AGN 
population-synthesis model. 
 


%\appendix

%% Generally speaking, only the figure captions, and not the figures
%% themselves, are included in electronic manuscript submissions.
%% Use \figcaption to format your figure captions. They should begin on a
%% new page.

%\clearpage

%% No more than seven \figcaption commands are allowed per page,
%% so if you have more than seven captions, insert a \clearpage

%% There must be a \figcaption command for each legend. Key the text of the
%% legend and the optional \label in curly braces. If you wish, you may
%% include the name of the corresponding figure file in square brackets.
%% The label is for identification purposes only. It will not insert the
%% figures themselves into the document.
%% If you want to include your art in the paper, use \plotone.
%% Refer to the on-line documentation for details.

%% The following command ends your manuscript. LaTeX will ignore any text

\end{document}